\documentclass[aps,prb,showpacs,twocolumn,floats]{revtex4}
\usepackage{graphicx,psfig}
\usepackage{dcolumn}
\usepackage{bm}

\input{tcilatex}

\begin{document}

\title{Butterfly hysteresis curve is a signature of adiabatic Landau-Zener
transition }
\author{Mark Vogelsberger, D. A. Garanin and Rolf Schilling}
\affiliation{\mbox{Institut f\"ur Physik,
Johannes-Gutenberg-Universit\"at, D-55099 Mainz, Germany}}
\date{22 July 2005}

\begin{abstract}
We stress that the so-called butterfly hysteresis curves observed in
dynamical magnetization measurements on systems of low-spin magnetic
molecules such as V$_{15}$ and V$_{6}$ are a signature of adiabatic
Landau-Zener transitions rather than that of a phonon bottleneck. We
investigate the magnetization dynamics analytically with the help of a
simple relaxation theory in the basis of the adabatic energy levels of the
spin 1/2, to a qualitative accordance with experimental observations. In
particular, reversible behavior is found near zero field, the corresponding
susceptibility being bounded by the equilibrium and adiabatic
susceptibilities from below and above, respectively.
\end{abstract}
 \maketitle

%\pacs{75.50.Xx,76.60.Es, 75.45.+j}

Magnetic hysteresis curves in crystals of molecular magnets with an
effective spin $S=1/2$ such as V$_{15}$ (Refs.\
\onlinecite
{chietal00prl,chietal03prbrc}) and V$_{6}$ (Ref.\ \onlinecite{rouetal05prl})
have under some conditions the so-called butterfly form that is
conventionally considered as a signature of the phonon bottleneck.\cite
{abrble70} Similar phenomenon has been observed on ferric wheels NaFe$_{6}$,
where the ground state changes from $S=0$ to $S=1$ at some magnetic field.
\cite{waletal02prl}

We shall demonstrate that butterfly hysteresis curves can qualitatively be
explained already by a simple relaxational model taking into account
adiabatic Landau-Zener transitions at avoided level crossing. Using the
additional equation for nonequilibrium phonons\cite{abrble70} does not
change the butterfly hysteresis curves qualitatively. While bad thermal
contact between the crystal and the holder can apparently lead to the
bottleneck and thus to smaller experimentally observed effective relaxation
rates for the spin, these rates could be reduced for other reasons, too. In
this case butterfly hysteresis curves arise without the bottleneck. Although
this fact is known (see, e.g., Ref.\ \onlinecite{chietal00prl}), an
appropriate quantitative theory is still lacking.

\begin{figure}[t]
\unitlength1cm
\begin{picture}(11,6)
\centerline{\psfig{file=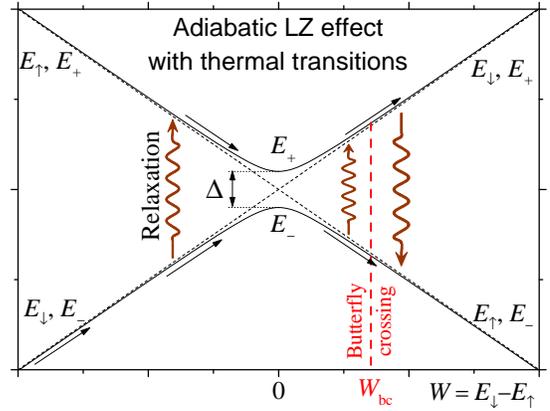,angle=-90,width=9cm}}
\end{picture}
\caption{Adiabatic Landau-Zener effect with thermal transitions. The
direction of relaxation changes at the energy bias $W_{bc}$ that corresponds
to the butterfly crossing of the magnetization curve, $m_{z}(t)=m_{z}^{(%
\mathrm{eq})}(t)$. }
\label{Fig-LZEffect}
\end{figure}

In V$_{15}$ the zero-field splitting $\Delta $ between the two low-lying
spin levels is about $\Delta /k_{B}\simeq $50$\div $80 mK (Refs.\
\onlinecite
{chietal00prl,chietal03prbrc}), and for the experimental sweep rate $%
dB_{z}/dt=0.1$ T/s the Landau-Zener parameter $\varepsilon =\pi \Delta
^{2}/(2\hbar v)$ with $v=2Sg\mu _{B}dB_{z}/dt$ is about 10$^{9}.$ This means
that, in the absence of relaxation, the system follows the lowest adiabatic
energy level $E_{-}$ (see Fig.\ \ref{Fig-LZEffect}), the probability to
remain in the original spin-down state $P=e^{-\varepsilon }$ being
negligibly small.

In the opposite limit of very fast relaxation between the adiabatic levels $%
E_{-}$ and $E_{+}$ the system relaxes up the energy, $E_{-}$ $\rightarrow
E_{+},$ before crossing the resonance and down the energy, $E_{+}$ $%
\rightarrow E_{-},$ after crossing the resonance, the level populations $%
n_{\pm }$ satisfying the equilibrium condition
\begin{equation}
n_{-}^{(\mathrm{eq})}-n_{+}^{(\mathrm{eq})}=\mathrm{tanh}\left( \frac{\hbar
\omega _{0}}{2k_{B}T}\right) ,  \label{nDiffEq}
\end{equation}
where
\begin{equation}
\hbar \omega _{0}\equiv E_{+}-E_{-}=\sqrt{W^{2}+\Delta ^{2}}
\label{homega0Def}
\end{equation}
and $W=2Sg\mu _{B}B_{z}$ is the energy bias that in the field-sweep
experiments is a linear function of time: $W=vt=2Sg\mu _{B}\left(
dB_{z}/dt\right) t$.

Changing to the spin-up/down basis allows to calculate the reduced
magnetization
\begin{equation}
m_{z}=\frac{W}{\sqrt{W^{2}+\Delta ^{2}}}\left( n_{-}-n_{+}\right) ,
\label{mzGeneral}
\end{equation}
where $n_{-}=1$ and $n_{+}=0$ far before crossing the resonance. In the
absence of relaxation (adiabatic case) one has $n_{-}\cong 1$ and $%
n_{+}\cong 0$ at any moment of time, while at equilibrium $n_{\pm }$ satify
Eq.\ (\ref{nDiffEq}) and the magnetization is given by
\begin{equation}
m_{z}^{(\mathrm{eq})}=\frac{W}{\sqrt{W^{2}+\Delta ^{2}}}\mathrm{tanh}\left(
\frac{\sqrt{W^{2}+\Delta ^{2}}}{2k_{B}T}\right) .  \label{mzEqui}
\end{equation}
In the case $k_{B}T\gg \Delta $ the equilibrium magnetization simplifies to $%
m_{z}^{(\mathrm{eq})}\cong \mathrm{tanh}\left[ W/(2k_{B}T)\right] ,$
independently of $\Delta .$ Note that, whatever the relaxation processes,
one has $m_{z}=0$ at resonance, $W=0.$

In the intermediate-relaxation regime the time-dependent values of $n_{\pm }$
can be obtained from the master equation that can be written in the form
\begin{equation}
\frac{d}{dt}\left( n_{-}-n_{+}\right) =-\Gamma \left[ n_{-}-n_{+}-\mathrm{%
tanh}\left( \frac{\hbar \omega _{0}}{2k_{B}T}\right) \right] .
\label{MasterEq}
\end{equation}
Note that in the adiabatic limit $\varepsilon \gg 1$ quantum-mechanical
transitions between the levels $E_{\pm }$ can be neglected, and there is no
need to use the full density-matrix equation. Using the adiabatic basis in
Eq.\ (\ref{MasterEq}) allows to describe the whole range of the field sweep $%
W,$ in contrast to the master equation in the basis of the eigenstates of $%
S_{z}$ of Ref.\ \onlinecite{rouetal05prl} that is applicable for $|W|\gg
\Delta $ only.

\begin{figure}[t]
\unitlength1cm
\begin{picture}(11,6)
\centerline{\psfig{file=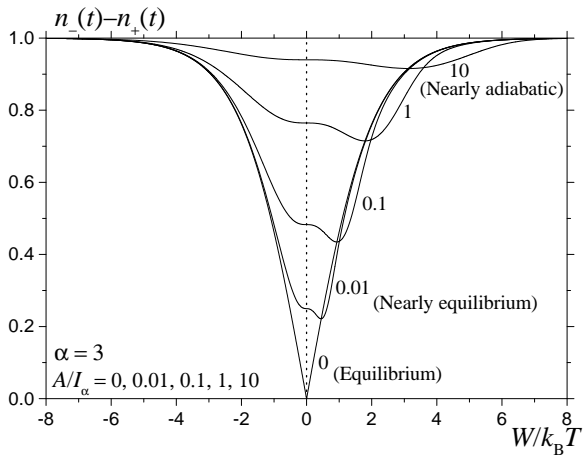,angle=-90,width=9cm}}
\end{picture}
\caption{The population difference $n_{-}(t)-n_{+}(t)$ obtained by
numerical integration in Eq.\ (\protect\ref{MasterEqSol}) vs the
energy bias $W$ for the linear sweep from $t=-\infty $ to $\infty
$ in the limit $\Delta /k_{B}T\rightarrow 0$ for different values
of the reduced sweep rate $A$ of Eq.\ (\protect\ref{ADefDef}) .}
\label{Fig-populations}
\end{figure}

Whereas the equilibrium solution of Eq.\ (\ref{MasterEq}) for $n_{-}-n_{+}$
that is attained in the fast-relaxation limit is even in $W$ with a minimum
at $W=0,$ the general nonequilibrium solution for $n_{-}-n_{+}$ is lagging
behind $n_{-}^{(\mathrm{eq})}-n_{+}^{(\mathrm{eq})}$ (see Fig.\ \ref
{Fig-populations})$.$ For crossing the resonance in the positive direction,
it has a minimum at some $W_{bc}>0.$ For $W<W_{bc}$ one has $%
n_{-}-n_{+}>n_{-}^{(\mathrm{eq})}-n_{+}^{(\mathrm{eq})},$ whereas for $%
W>W_{bc}$ one has $n_{-}-n_{+}<n_{-}^{(\mathrm{eq})}-n_{+}^{(\mathrm{eq})}.$
This overshoot leads to the butterfly hysteresis curves for $m_{z}(t)$
defined by Eq.\ (\ref{mzGeneral}): At $W=W_{bc}$ the dynamic magnetization $%
m_{z}(t)$ crosses the equilibrium magnetization curve,
\begin{equation}
m_{z}(t_{bc})=m_{z}^{(\mathrm{eq})}(t_{bc}),  \label{ButtCrossDef}
\end{equation}
that is the definition of the butterfly crossing$.$

The relaxation rate $\Gamma $ in Eq.\ (\ref{MasterEq}) is mainly due to the
direct phonon processes and, in the absence of the phonon bottleneck, it has
the form
\begin{equation}
\Gamma =\Gamma _{0}\mathrm{coth}\left( \frac{\hbar \omega _{0}}{2k_{B}T}%
\right) .  \label{GammviaGamma0}
\end{equation}
Here $\Gamma _{0}$ depends on the details of the spin-phonon coupling (for V$%
_{15}$ and V$_{6}$ one of the candidates is Dzialoshinskii-Moriya
interaction \cite{chietal03prbrc}). We are not going to discuss the details
of the spin-phonon interactions here as the butterfly curve is quite a
general phenomenon. One has only to take into account that $\Gamma $ depends
on time via the transition frequency $\omega _{0}.$ We set
\begin{equation}
\Gamma _{0}=\Gamma _{00}\left( \frac{\hbar \omega _{0}}{\Delta }\right)
^{\alpha },  \label{Gamma00Def}
\end{equation}
where in most cases $\alpha =3.$ For a better fit to experiments, one can
include an $\omega _{0}$-independent relaxation rate in Eq.\ (\ref
{GammviaGamma0}), as was done in Ref.\ \onlinecite{rouetal05prl}. The
solution of Eq.\ (\ref{MasterEq}) has the form
\begin{equation}
n_{-}(t)-n_{+}(t)=e^{-\widetilde{\Gamma }t}\int_{-\infty }^{t}dt^{\prime
}\Gamma (t^{\prime })e^{\widetilde{\Gamma }(t^{\prime })t^{\prime }}\mathrm{%
tanh}\left( \frac{\hbar \omega _{0}(t^{\prime })}{2k_{B}T}\right) ,
\label{MasterEqSol}
\end{equation}
where
\begin{equation}
\widetilde{\Gamma }(t)\equiv \frac{1}{t}\int_{0}^{t}dt^{\prime }\Gamma
(t^{\prime }).  \label{GammaTilDef}
\end{equation}
it is easy to check that $\mathrm{lim}_{t\rightarrow \pm \infty }\left[
n_{-}(t)-n_{+}(t)\right] =1.$ The strongest deviation of $n_{-}(t)-n_{+}(t)$
from 1 originates from such $t$ for which $\mathrm{tanh}\left[ \hbar \omega
_{0}(t^{\prime })/(2k_{B}T)\right] $ in Eq.\ (\ref{MasterEqSol}) is small
enough compared to one. It is the range where thermal transitions become
significant.

\begin{figure}[t]
\unitlength1cm
\begin{picture}(11,6)
\centerline{\psfig{file=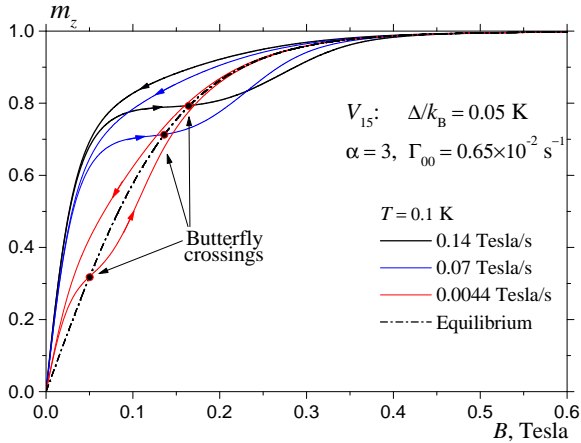,angle=-90,width=9cm}}
\end{picture}
\caption{Magnetization hysteresis curves $m_{z}(t)$ calculated from Eqs.\ (%
\protect\ref{mzGeneral}) and (\protect\ref{mzEqui}) for the
parameters of Ref.\ \onlinecite{chietal00prl}.}
\label{Fig-magnetization}
\end{figure}

In the case $\Delta \sim k_{B}T$ that was realized in experiments on V$_{15}$
in Refs.\ \onlinecite{chietal00prl,chietal03prbrc} ($\Delta /k_{B}\simeq
0.05 $ K, $T=0.1\div 0.2$ K) the thermal transitions are important in the
vicinity of the level crossing: $W=vt\sim \Delta $. As the system spends the
time $t_{th}\sim \Delta /v\sim k_{B}T/v$ in this region, the case $\Gamma
_{00}\ll v/\Delta \sim v/(k_{B}T)$ corresponds to the adiabatic limit, $%
n_{-}\cong 1$ and $n_{+}\cong 0,$ whereas the case $\Gamma _{00}\gg v/\Delta
\sim v/(k_{B}T)$ corresponds to the equilibrium situation, $%
n_{-}(t)-n_{+}(t)\cong n_{-}^{(\mathrm{eq})}(t)-n_{+}^{(\mathrm{eq})}(t).$
In the intermediate case $\Gamma _{00}\sim v/\Delta \sim v/(k_{B}T)$ the
dynamical relaxation effect is mostly pronounced, and the hysteresis curve
has a butterfly shape with the crossing with the equilibrium magnetization
curve at $W_{bc}\sim \Delta \sim k_{B}T$. Our numerical results obtained
from Eq.\ (\ref{MasterEqSol}) and shown in Fig.\ \ref{Fig-magnetization})
are in a qualitative agreement with experimental and theoretical results of
Ref.\ \onlinecite
{chietal00prl}, except for the slowest-sweep curve $dH_{z}/dt=0.0044$ T/s.

The low-temperature case $\Delta \gg k_{B}T$ is trivial, since here there
are no thermal transitions and the adiabatic solution $n_{-}\cong 1$ and $%
n_{+}\cong 0$ is valid for all times, while the magnetization follows from
Eq.\ (\ref{mzGeneral}).

In the high-temperature range $\Delta \ll k_{B}T$ (in experiments on V$_{6}$
in Ref.\ \onlinecite{rouetal05prl} the estimated energy gap is $\Delta
/k_{B}\simeq 0.4$ K, whereas the maximal temperature was $T=4.2$ K) thermal
transitions take place in a much broader region $\left| W\right| \sim
W_{0}=k_{B}T$ than the quantum-mechanical crossing $\left| W\right| \sim
\Delta .$ The system spends the time
\begin{equation}
t_{th}=k_{B}T/v  \label{tcDef}
\end{equation}
in the thermal-transition range. Neglecting the narrow region $\left|
W\right| \sim \Delta $ and approximating $\hbar \omega _{0}\cong \left|
W\right| =v\left| t\right| $ one obtains from Eqs.\ (\ref{GammaTilDef}) and (%
\ref{Gamma00Def})
\begin{equation}
\widetilde{\Gamma }(t)t\cong \frac{\Gamma _{00}k_{B}T}{v}\left( \frac{k_{B}T%
}{\Delta }\right) ^{\alpha }\Phi _{\alpha }\left( \frac{v\left| t\right| }{%
k_{B}T}\right) \mathrm{sign}(t),  \label{GammaTilThermal}
\end{equation}
where
\begin{equation}
\Phi _{\alpha }\left( z\right) \equiv \int_{0}^{z}du\,u^{\alpha }\mathrm{coth%
}\left( \frac{u}{2}\right) .  \label{PhialzDef}
\end{equation}
The most pronounced dynamical relaxation effect leading to the butterfly
hysteresis curve can be expected to take place for
\begin{equation}
\widetilde{\Gamma }(t_{th})t_{th}\equiv \frac{1}{A}\sim 1,  \label{ADef}
\end{equation}
where we have introduced
\begin{equation}
A=\frac{v}{\Gamma _{00}k_{B}T}\left( \frac{\Delta }{k_{B}T}\right) ^{\alpha }%
\frac{1}{\Phi _{\alpha }\left( 1\right) }.  \label{ADefDef}
\end{equation}
as a reduced relaxation time or reduced sweep rate and $\Phi _{1}\left(
1\right) =2.06,$ $\Phi _{2}\left( 1\right) =1.04,$ $\Phi _{3}\left( 1\right)
=0.670,$ $\Phi _{4}\left( 1\right) =0.527,$ $\Phi _{5}\left( 1\right) =0.424$%
.

\begin{figure}[t]
\unitlength1cm
\begin{picture}(11,6)
\centerline{\psfig{file=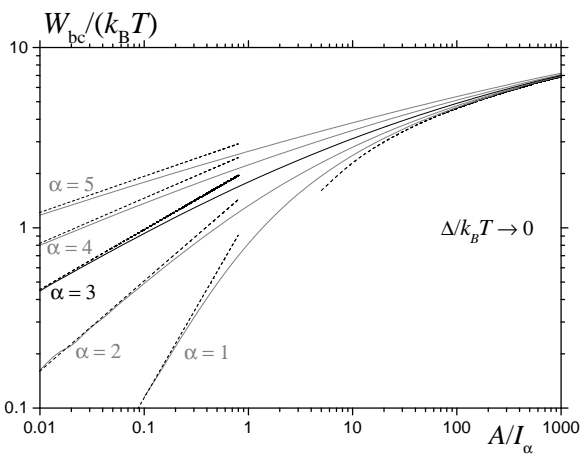,angle=-90,width=9cm}}
\end{picture}
\caption{Butterfly-crossing bias $W_{bc}$ vs the reduced
relaxation time $A$ for $\protect\alpha =1..5.$ The
nearly-equilibrium asymptotes of Eq.\ (\protect\ref
{WcQuasiEqu}) on the left side and the nearly-adiabatic asymptote of Eq.\ (%
\protect\ref{WcQuasiAdiabatic}) on the right side are shown by the
dashed lines.} \label{Fig-WcvsA}
\end{figure}

One can find the crossing point $W_{bc}$ of the butterfly hysteresis curve $%
m_{z}(t)=m_{z}^{(\mathrm{eq})}(t)$ from Eq.\ (\ref{MasterEqSol}) by
requiring $n_{-}(t)-n_{+}(t)=n_{-}^{(\mathrm{eq})}(t)-n_{+}^{(\mathrm{eq}%
)}(t).$ For $\Delta \ll k_{B}T$ in the nearly adiabatic limit $A\gg 1$ one
obtains (for the increasing field)
\begin{equation}
W_{bc}=vt_{bc}\cong k_{B}T\mathrm{ln}\left( \frac{A}{I_{\alpha }}\right)
\label{WcQuasiAdiabatic}
\end{equation}
and
\begin{equation}
n_{-}(t_{bc})-n_{+}(t_{bc})=\mathrm{tanh}\frac{W_{bc}}{2k_{B}T}\cong 1-\frac{%
2I_{\alpha }}{A},  \label{ncQuasiAdiabatic}
\end{equation}
where
\begin{equation}
I_{\alpha }\equiv \int_{0}^{\infty }dz\frac{z^{\alpha }}{\Phi _{\alpha
}\left( 1\right) }\left[ \mathrm{coth}\left( \frac{z}{2}\right) -1\right]
\label{IalphaDef}
\end{equation}
$I_{1}=1.60,$ $I_{2}=4.62,$ $I_{3}=18.6,$ $I_{4}=94.4,$ $I_{5}=576.$ One can
see that, in fact, the apparent applicability condition of Eqs.\ (\ref
{WcQuasiAdiabatic}) and (\ref{ncQuasiAdiabatic})\ is $I_{\alpha }/A\ll 1$
that results in rather large values of $A$ for $\alpha =3$ and higher. This
is an essential correction to the \textit{a priori} estimation of Eq.\ (\ref
{ADef}). With increasing $\alpha $ the validity range of Eqs.\ (\ref
{WcQuasiAdiabatic}) and (\ref{ncQuasiAdiabatic}) becomes narrower than $%
I_{\alpha }/A\ll 1$ since the former is the first term of the asymptotic
expansion in powers of $1/A$ that becomes progressively bad with increasing $%
\alpha .$ The physical origin of this difficulty is the following. For large
powers $\alpha $ the situation becomes inhomogeneous in the whole domain of $%
W$: It is nearly adiabatic for smaller $W$ and nearly equilibrium for larger
$W.$

For $\Delta \ll k_{B}T$ in the nearly equilibrium limit $A\ll 1$ one obtains
\begin{equation}
W_{bc}\cong k_{B}T\left( q_{\alpha }A\right) ^{1/\alpha }  \label{WcQuasiEqu}
\end{equation}
and
\begin{equation}
n_{-}(t_{bc})-n_{+}(t_{bc})\cong \frac{1}{2}\left( q_{\alpha }A\right)
^{1/\alpha },  \label{ncQuasiEqu}
\end{equation}
where $q_{\alpha }\equiv \left( \alpha /2\right) \Phi _{\alpha }\left(
1\right) p_{\alpha }$ and $p_{\alpha }$ is the solution of the transcedental
equation
\begin{equation}
p_{\alpha }^{1/\alpha }=e^{-p_{\alpha }}\Gamma \left( 1+\frac{1}{\alpha }%
\right) +\int_{0}^{p_{\alpha }}dp\,p^{1/\alpha }e^{p-p_{\alpha }}.
\label{palphaEq}
\end{equation}
One obtains $p_{1}=0.693,$ $p_{2}=0.535,$ $p_{3}=0.482,$ $p_{4}=0.455,$ $%
p_{5}=0.438,$ and $q_{1}=0.712,$ $q_{2}=0.557,$ $q_{3}=0.506,$ $q_{4}=0.480,$
$q_{4}=0.464.$ Note that the general applicability condition of our
approximation replacing $\hbar \omega _{0}\Rightarrow \left| W\right| $ is $%
W_{bc}\gg \Delta .$ This yields the applicability condition for Eqs.\ (\ref
{WcQuasiEqu}) and (\ref{ncQuasiEqu})
\begin{equation}
\left( \frac{\Delta }{k_{B}T}\right) ^{\alpha }\ll A\ll 1.
\label{AQuasiEquApplic}
\end{equation}

The full dependence of $W_{bc}$ on the reduced relaxation time $A$ of Eq.\ (%
\ref{ADefDef}) can be found numerically from Eq.\ (\ref{MasterEqSol}). The
results for $W_{bc}/(k_{B}T)$ for different $\alpha $ in the limit $\Delta
\ll k_{B}T$ are shown in Fig.\ \ref{Fig-WcvsA}. Note that at the butterfly
crossing the difference $n_{-}(t)-n_{+}(t)$ attains a minimum, according to
Eq.\ (\ref{MasterEq}). The magnetization value at the butterfly crossing $%
m_{z}^{(\mathrm{eq})}(t_{bc})$ can be obtained from Eq.\ (\ref{mzEqui}) and
it is given by
\begin{equation}
m_{z}(t_{bc})=\frac{W_{bc}}{\sqrt{W_{bc}^{2}+\Delta ^{2}}}\mathrm{tanh}\frac{%
\sqrt{W_{bc}^{2}+\Delta ^{2}}}{2k_{B}T}.  \label{mzbc}
\end{equation}

\begin{figure}[t]
\unitlength1cm
\begin{picture}(11,6)
\centerline{\psfig{file=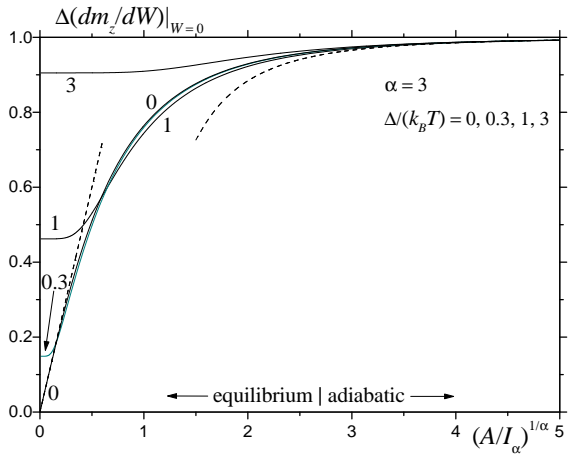,angle=-90,width=9cm}}
\end{picture}
\caption{Susceptibility at the level crossing vs $A$ for $\protect\alpha =3$
and different values of $\Delta /(k_{B}T).$ The asymptotes given by Eqs.\ (%
\protect\ref{chiLargeA}) and (\protect\ref{chiSmallA}) are shown
by the dashed lines.} \label{Fig-sus}
\end{figure}

Let us now consider the reduced magnetic susceptibility $\tilde{\chi}=\Delta
(dm_{z}/dW)$ at the level crossing, $W=0.$ Evidently $\tilde{\chi}$ depends
on the sweep rate and is bounded by its equilibrium and adiabatic values:
\begin{equation}
\tilde{\chi}_{\mathrm{eq}}=\mathrm{tanh}\frac{\Delta }{2k_{B}T}\leq \tilde{%
\chi}\leq 1=\tilde{\chi}_{\mathrm{ad}}.  \label{chibounds}
\end{equation}
With the help of Eq.\ (\ref{mzGeneral}) one obtains the result
\begin{equation}
\tilde{\chi}=n_{-}(0)-n_{+}(0).  \label{chivianpm}
\end{equation}
It is easy to prove that $n_{-}(0)-n_{+}(0)$ is the same for the increasing
field and decreasing field. Then, Eq.\ (\ref{chivianpm}) implies that the
zero-field susceptibility $\tilde{\chi}$ does not depend of the direction of
the field. Accordingly the magnetization, although different from the
equilibrium one, is reversible near zero field, in full agreement with the
experimental finding.\cite{chietal00prl} Note that the theoretical aproach
of Ref.\ \onlinecite{rouetal05prl} is not applicable in the vicinity of the
Landau-Zener crossing. Using Eqs.\ (\ref{MasterEqSol}) and (\ref{chivianpm})
one can express $\tilde{\chi}$ as
\begin{equation}
\tilde{\chi}=\int_{-\infty }^{0}dt\Gamma (t)e^{\widetilde{\Gamma }(t)t}%
\mathrm{tanh}\left( \frac{\hbar \omega _{0}(t)}{2k_{B}T}\right)
\label{chiGenRes}
\end{equation}
that is plotted vs $A/I_{\alpha }$ for different values of $\Delta /(k_{B}T)$
in Fig.\ \ref{Fig-sus}. In the nearly adiabatic limit $A\gg 1$ the asymptote
of Eq.\ (\ref{chiGenRes}) is
\begin{equation}
\tilde{\chi}\cong 1-\frac{2\Gamma (\alpha +1)}{\Phi _{\alpha }\left(
1\right) A}.  \label{chiLargeA}
\end{equation}
In the nearly equilibrium limit $A\ll 1,$ the most interesting is the
asymptote corresponding to $\Delta /(k_{B}T)\rightarrow 0$:
\begin{equation}
\tilde{\chi}\cong \left( \alpha \Phi _{\alpha }\left( 1\right) A\right)
^{1/\alpha }2^{-(1+1/\alpha )}\Gamma \left( 1+1/\alpha \right) .
\label{chiSmallA}
\end{equation}

Summarizing, we gave a detailed analytical consideration of the adiabatic
Landau-Zener effect with relaxation and we have shown that this is a minimal
model to describe experimentally observed butterfly hysteresis curves. More
complicated relaxation models \cite{abrble70} including an additional
kinetic equation for nonequilibrium phonons in the case of the phonon
bottleneck seem to be nonessential to describe the butterfly hysteresis
curves. One can interpret the latter in the framework of the simplest
relaxation theory described above by choosing an appropriate relaxation rate
$\Gamma $ that can be effectively reduced because of a poor thermal contact
of the crystal with the holder. If the details of the phonon dynamics
related to the phonon bottleneck are the subject of investigation, a special
care should be taken to single out their effect on the magnetization
hysteresis curves.

\bibliographystyle{apsrev}
\bibliography{gar-oldworks,gar-books,gar-own,gar-tunneling,gar-relaxation}

\end{document}